\documentclass[aps,twocolumn,showpacs]{revtex4}
\usepackage{graphicx}
\usepackage{epsf}
\usepackage{amsmath}
\usepackage{textcomp}
\usepackage{amssymb}

\begin{document}
\title{Shock waves in thermal lensing}
\vspace{0.5cm}

\author{S. Gentilini$^{1\star}$, N. Ghofraniha$^2$, E. DelRe$^{2,3}$ and C. Conti$^{1,3}$}
\affiliation{\small
$^1$ ISC-CNR c/o Dipartimento di Fisica - Universit\`{a} di Roma La Sapienza, P. A. Moro
2, 00185, Rome, Italy\\
{$^2$ IPCF-CNR c/o Dipartimento di Fisica - Universit\`{a} di Roma La Sapienza, P. A. Moro
2, 00185, Rome, Italy\\
$^3$ Dipartimento di Fisica - Universit\`{a} di Roma La Sapienza, P. A. Moro
2, 00185, Rome, Italy\\
$^*$Corresponding author: silvia.gentilini@roma1.infn.it
} }

\date{\today}

\begin{abstract}

We review recent experimental investigation on spatial shock waves formed by the
self-defocusing action of a laser beam propagating in disordered thermal nonlinear media.

\end{abstract}


\maketitle

\section{Introduction}

Dispersive shock waves (DSWs) are observed in nonlinear optics in systems described by the nonlinear Sch\"{o}rdinger equation, when the so-called hydrodynamical reduction is valid \cite{gurevich,bronskibook,kamchatnov,conti09,whitman,Besse12}. The introduction of a small amount of disorder competes with nonlinearity and hampers the shock formation \cite{Ghoprl12,Gentilini12}. This makes the DSWs an appealing framework to study the interplay between randomness and nonlinear waves, a subject of growing interest as witnessed by recent theoretical \cite{Con11,Fol12,kivshar10} and experimental studies \cite{DelRe11,levi11,Fleisher2012}.

At variance with the ordered systems \cite{wan07}, the direct observation and characterization of optical shock waves in the presence of structural randomness is burdened by several technical difficulties in identifying an appropriate nonlinear medium, feasible excitation conditions, and relevant observables. There are several possibilities to characterize the excitation of \emph{undular bores} and related phenomena \cite{deykoon, wyller, wurtz,ell,hoefer,barsi:07}: the very definition and observation of the wave-breaking phenomena in the presence of disorder and nonlinearity is an open issue. This calls for an extensive development of experimental techniques and the use of multiple methods to characterize the DSWs.

In this paper we give a detailed review of our experimental investigations of the hydrodynamical regime in the generation of optical shocks during nonlinear optical propagation in a thermal defocusing nonlinearity of a continuous wave (CW) laser beam. The hydrodynamical regime is achieved when the nonlinear length is much smaller than the diffraction and losses (absorption and scattering) lengths.

Our experimental technique allows the direct observation of a propagating initially Gaussian laser beam in a thermal nonlinear liquid with controllable disorder obtained by a colloidal dispersion with a low index of refraction contrast. We show that, by increasing the strength of the nonlinearity, shock formation is enhanced, while, on the other hand, the random scatterers limit and ultimately inhibit the wave-breaking phenomenon.
We quantify such a competition by analyzing both the laser beam along the propagation direction and its far field distribution intensity: these allow the measurement of the relevant scaling laws that relate the shock position \cite{Gho12,Ghoprl12} and the post-shock wave-vector spectrum with the input beam size, power, and the scale and strength of disorder \cite{Gentilini12}.

These observables, namely the shock formation point and the output wave-vector spectrum, exhibit a threshold and determine a phase diagram identifying parameter regions where the shock occurs.

The paper is organized as follows: first we briefly review the theoretical framework. We then illustrate the experimental setup and the characterization of the samples used in the experiments. We then review the results in three different sections, corresponding to the experimental characterization of different observables. We present the analysis of the beam intensity distribution both during propagation and exiting the samples. We conclude with a summary section of the obtained results.

\section{Theoretical framework}\label{section1}
In our experiments, CW laser beams propagate in dye-doped dispersions of dielectric colloidal beads. The beam is partially absorbed and scattered, activating the interplay between thermal-defocusing and spatial-disorder.

Neglecting, in a first approximation, the spatial nonlocality \cite{ghofraniha}, the refractive index perturbation in the presence of nonlinearity and disorder to the bulk index $n_0$ is written as:
\begin{equation}
\Delta n=n_2I+\Delta n_R(X,Y,Z)
\label{eq:refractiveindex}
\end{equation}
where $n_2<0$ takes into account the considered defocusing Kerr effect, $I$ is the optical intensity and $\Delta n_R$ represents the random perturbation due to the colloidal beads.

The propagation of a TEM$_{00}$ Gaussian beam inside the medium is described by the paraxial wave equation for the complex envelope, $A$, of a monochromatic electric field $E=(\frac{2}{c\epsilon_0n_0})^{1/2}A\exp({ikZ-i\omega T})$,
\begin{equation}
2ik\frac{\partial A}{\partial Z}+\nabla^2_{X,Y}A+2k^2\frac{\Delta n}{n_0}A=0,
\label{eq:paraxial}
\end{equation}
where $k=2\pi n_0/\lambda$ is the wave-vector, $c$ the velocity of light, and $\epsilon_0$ is the electric permittivity of free space. In Eq. (\ref{eq:paraxial}) $A$ is normalized such that $I=|A|^2$. Indicating with $I_0$ the input peak intensity, $w_0$ the input beam waist, $L_{nl}=n_0/(k_0|n_2|I_0)$ the nonlinear length scale, and introducing the scaled coordinates $x,y,z=X/w_0,Y/w_0,Z/L$, and the normalized field $\psi=A/\sqrt{I_0}$, we obtain the following dimensionless equation:
\begin{equation}
i\epsilon\frac{\partial\psi}{\partial z}+\frac{\epsilon^2}{2}\nabla^2_{x,y}\psi-|\psi|^2\psi+U_R\psi=0,
\label{eq:dimensionless}
\end{equation}
where $\epsilon\equiv L_{nl}/L=\sqrt{L_{nl}/L_d}$, being $L\equiv\sqrt{L_{nl}L_d}$ and $L_d=kw_0^2$ the diffraction length, and $U_R=\Delta n_R/(n_2I_0)$. The quantity $\epsilon$ measures the strength of the nonlinearity with respect to the diffraction: a small value for $\epsilon$ implies negligible diffraction and a pronounced nonlinear response. $U_{R}$ is the ratio between the perturbation of index due to the disorder and the nonlinearity.

Setting in Eq.(\ref{eq:dimensionless}) $\psi=\sqrt{\rho(r,z)}\exp[i\phi(r,z)/\epsilon]$ and retaining only the leading order in $\epsilon$, we obtain the following equation for the phase $\phi$:
\begin{equation}
\phi_z+\frac{1}{2}(\phi_x^2+\phi_y^2)+\rho-U_R=0
\label{eq:decoupled}
\end{equation}

Limiting to one dimension ($1$D, $\partial_y=0$), performing the transverse derivative of Eq.(\ref{eq:decoupled}), and defining a \emph{velocity field} equal to the phase chirp, $u\equiv\phi_x$, we have
\begin{equation}
u_z+uu_x+\partial_x(\rho-U_R)=0.
\label{eq:hopf}
\end{equation}
In the homogeneous case ($\rho=$const.) and for an ordered medium ($U_R=0$), Eq. (\ref{eq:hopf}) takes the form of the Hopf equation \cite{whitman}, the solution of which can develop discontinuities in the velocity profile, $u_x\rightarrow\infty$, and hence gives rise to shock waves.

Here we remark that from the hydrodynamical approximation, a threshold in the nonlinearity is present:
in fact, the approximation holds true when $L_{nl}\ll L_d$.
Another threshold arises from the term $U_R=\Delta n_R/(n_2I_0)$, corresponding to the existence of a critical value for the amount of randomness, above which it is expected that no shock occurs: when the random index perturbation $\Delta n_R$ becomes comparable with the nonlinearity $n_2I_0$, the material refractive index fluctuations are so pronounced that the nonlinear effect is totally masked.

Correspondingly, in our experiments (see below) in absence of disorder, we find a threshold in the laser power, while in the presence of disorder a threshold emerges also in the amount of randomness.

\section{Experimental setup.}

In the hydrodynamic limit, DSWs are expected to occur
when the nonlinearity is dominant compared to diffraction; nevertheless the diffraction, which is initially negligible, starts to play a major role in the proximity of the
wave-breaking point, and regularizes the singularity by means of the appearance of characteristic oscillations (undular bores).

Besides these regularizing oscillations, the singularity in the field phase and amplitude also results in a diffraction enhancement, evident in the funnel shape along the propagation direction (see below) appearing with the increase of the input power. This shows that the shock involves the spatial spectrum of the beam as detected in far field measurements.

\begin{figure*}
\includegraphics[width=\textwidth]{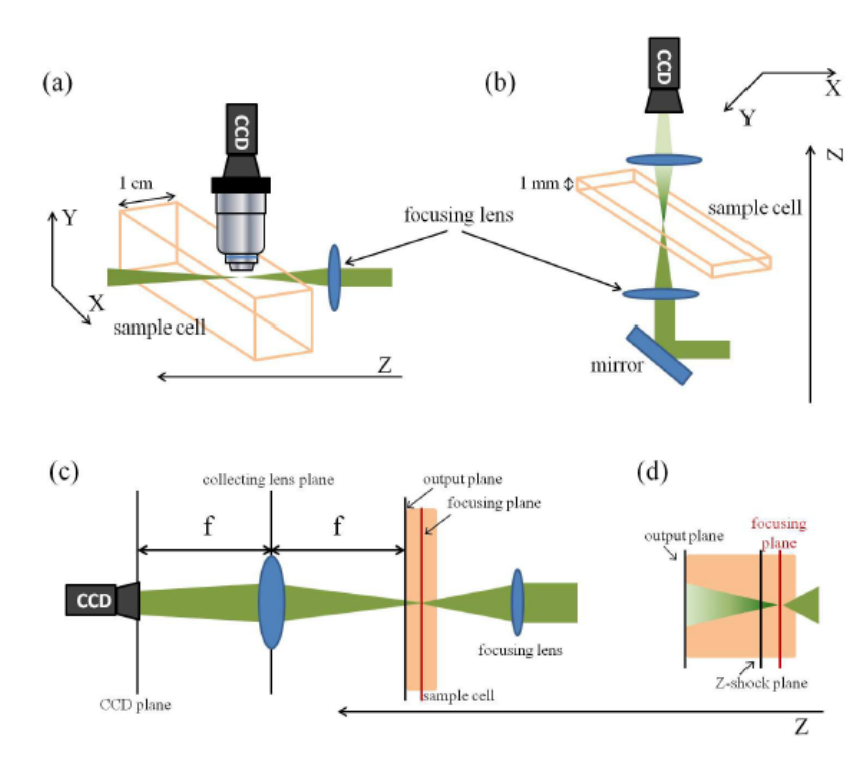}
\caption{(Color online) Experimental setup: (a) detection of the top fluorescence emission of the beam; (b) configuration for the collection of the far field intensity; (c) details of the optical setup of panel (b); (d) sketch showing the position of the shock plane inside of the sample.}\label{figexp1}
\end{figure*}

\emph{Near-field configuration} - In the near-field configuration our setup [Fig. \ref{figexp1}(a)] allows a direct visualization of the propagating beam profile, i.e., the intensity as function of the transverse coordinate, $X$, and of the propagation direction $Z$. This enables the identification of the shock point $Z_s$ as the propagation distance at which the maximum chirp occurs (see below).
Typically, a CW laser at wavelength $\lambda=532$nm is focused inside the sample. The beam waist in the focus is $w_0=10\mu$m.
The near-field configuration is sketched in Fig. \ref{figexp1}(a). A $1$cm $\times1$cm $\times3$cm glass cell is used and the laser beam propagates along the $1$ cm side. Top images of the fluorescence emission are collected by a MZ$16$ Leica microscope placed perpendicularly to the propagation direction, $Z$, and recorded by a $1024\times1392$ pixels CCD camera.

\emph{Far-field configuration} -
In Fig. \ref{figexp1}(b) we show the setup for the far-field measurements. The CW laser beam is focused inside the sample ($w_0=50\mu$m). The liquid samples are placed in a $1$mm $\times1$cm $\times3$cm glass cell, the laser beam propagates along the $1$mm side, the cell is placed in vertical direction in order to moderate the effect of heat convection.
As shown in Fig. \ref{figexp1}(c) the intensity distribution of the Fourier transform of the transmitted beam is collected by a CCD camera placed at the focal length from the collecting lens.
We calibrate the CCD detector by fitting with the Airy function the experimentally obtained Fourier transform of a $500\mu$m diameter pinhole, placed on the exit face of the cell. The angular spreading $\theta$ is related to the transverse wavevector as $k_{X,Y}=(2\pi/\lambda)\sin(\theta)$.

Figure \ref{figexp1}(d) shows the mutual positions of the focus plane, the shock plane and the output plane.

\section{Sample characterization}
As in previous experimental works, we use the thermal Kerr-like defocusing nonlinearity of
absorbing dye-doped liquid  media \cite{ghofraniha,Gho07,GhoLang,Gho12,Gentilini12,Ghoprl12}.
\begin{figure}[ht]
\includegraphics[width=8.5cm]{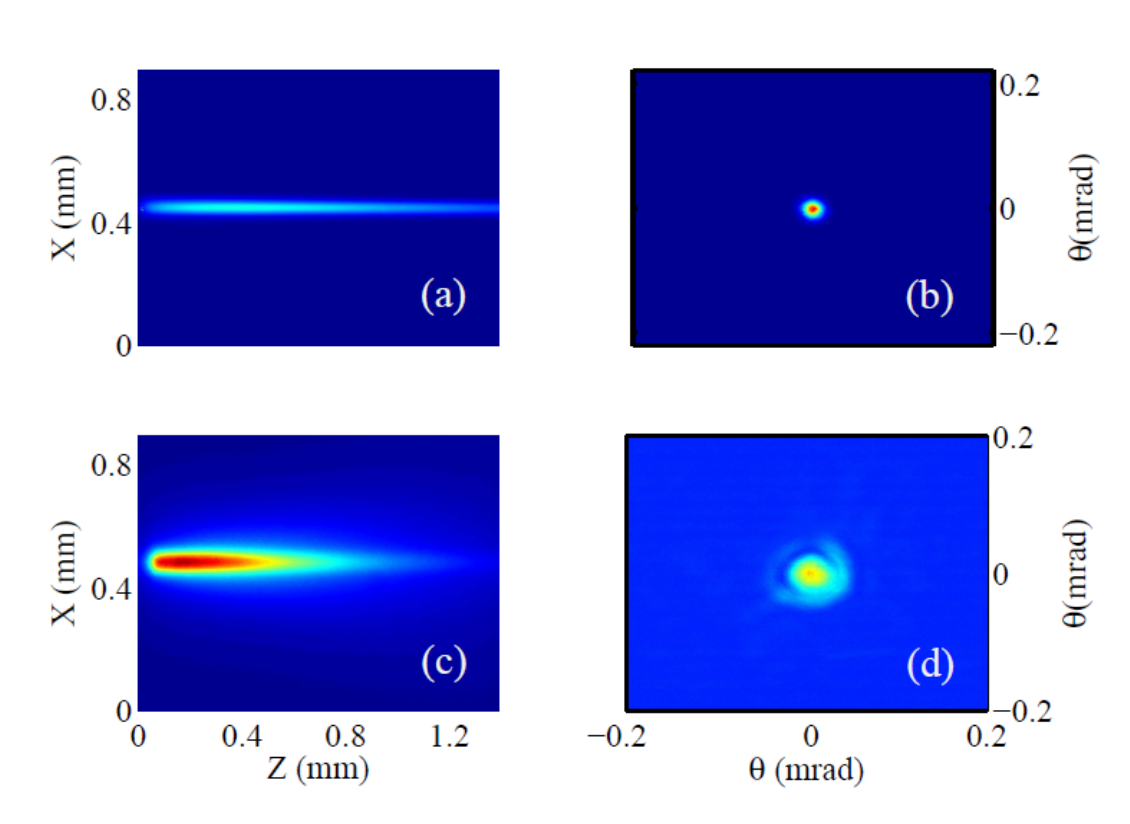}
\caption{(Color online) (a)--(c) Near-field images of the fluorescence emission of the propagating beam at fixed laser power ($P=8$mW) and two different particle concentrations, (a) $c_{SiO_2}=0$, (c) $c_{SiO_2}=0.03$w/w; (b)--(d) corresponding intensity profiles of the far-field images of the transmitted field at the exit face of $1$mm thick cell.
\label{figexp2}}
\end{figure}

\begin{figure}[ht]
\includegraphics[width=8.5cm]{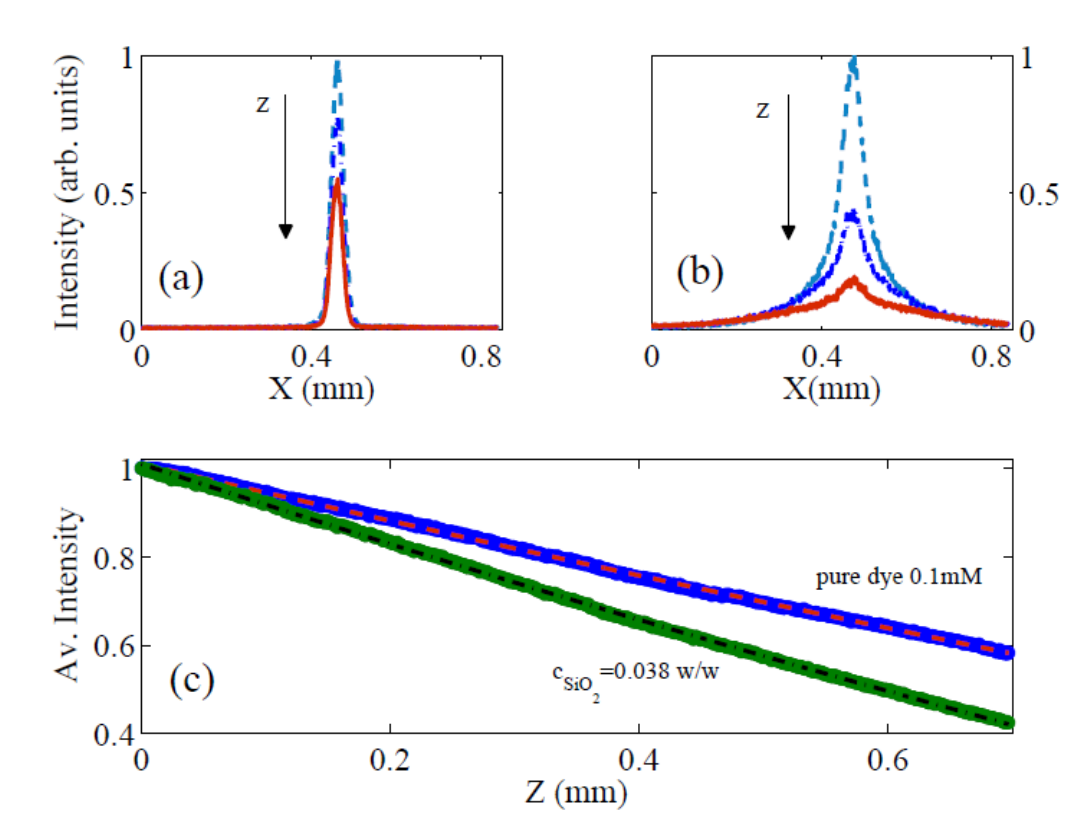}
\caption{(Color online) (a)--(b) Intensity profiles taken at three different $Z$ position in the images of Figs. \ref{figexp2}(a)--\ref{figexp2}(c): dashed line corresponds to $Z=0.3$mm, dot-dashed line to $Z=0.7$mm, and the continuous line to $Z=1.0$mm. (c) Exponential (quasi-linear) decays of beam intensity calculated from images of Fig. \ref{figexp2}(a) (dashed line) and Fig. \ref{figexp2}(c) (dot-dashed line).
\label{figexp3}}
\end{figure}

Our samples are aqueous solutions of Rhodamine B (RhB). We tailor the degree of absorption and nonlinearity by varying the concentration of RhB (c$_{RhB}$) from $0.05$ to $0.2$mM. We add disorder by using monodisperse $1\mu$m diameter silica (SiO$_2$) spheres. The degree of randomness is fixed by varying the concentration of SiO$_2$ (c$_{SiO_2}$) from $0.005$ to $0.04$w/w, in units of weight of silica particles over suspension weight. In terms of refractive index perturbation, the amount of disorder can be estimated by the following relation:
\begin{equation}
\langle\Delta n_R^2\rangle^{1/2}=c_{SiO_2}\rho_{H_2O}(n_{SiO_2}-n_{H_2O})/\rho_{SiO_2},
\label{eq:refractive_disorder}
\end{equation}
being n$_{SiO_2}$ (n$_{H_2O}$) and $\rho_{SiO_2}$ ($\rho_{H_2O}$) the refractive index and the density of the SiO$_2$ (H$_2$O), respectively.
The angular brackets in Eq. (\ref{eq:refractive_disorder}) denotes volume average.
Being the silica (water) density $\rho_{SiO_2}=2$g/cm$^3$ ($\rho_{H_2O}\approx 1$g/cm$^3$ at $25$\textcelsius), for the considered range of $c_{SiO_2}$ concentration, $\langle\Delta n^2_R\rangle^{1/2}$ varies between $4\times10^{-4}$ and $32\times10^{-4}$. Therefore, since from the theory a threshold in the disorder amount is predicted when $\langle\Delta n^2_R\rangle^{1/2}$ becomes comparable with the nonlinear perturbation  $|n_2|I_0\cong10^{-3}$, such a threshold is expected for the silica concentration $c_{SiO_2}=0.030$w/w as it was confirmed by our experiments (see below).

In our samples there are two leading loss mechanisms: (ii) absorption due to the RhB dye, and (ii) scattering due to SiO$_2$ particles. We find that scattering losses are predominant; this is shown in Fig. \ref{figexp2} where we compare the images of the transverse beam intensity distribution versus the propagation direction $Z$ at two different SiO$_2$ concentrations, fixed laser power $P$ and RhB concentration $c_{RhB}$.

Figure \ref{figexp2}(a) shows the top fluorescence of the laser beam in a pure dye sample ($c_{SiO_2}=0$), Fig. \ref{figexp3}(b) gives the corresponding far-field. Figures \ref{figexp2} (c) and \ref{figexp2}(d) report the case of a silica-dye sample at $c_{SiO_2}=0.03$w/w; the beam is more diffused and the far field reveals an enhanced spectral content. In the presence of disorder ($c_{SiO_2}>0$) the transverse spread of the beam along $Z$ is enhanced. This is clarified in the analysis reported in Fig. \ref{figexp3}(a) and \ref{figexp3}(b), which show the intensity profiles at three different $Z$ positions for the $0.1$mM pure dye solution and dye solution with silica at $c_{SiO_2}=0.03$w/w concentration, respectively. At variance with linear absorption, scattering due to SiO$_2$ beads, broadens the beam. Figure \ref{figexp3}(c) shows the average intensity Vs Z calculated from the images of Figs. \ref{figexp2}(a) and \ref{figexp2}(c) and the exponential decays that fit the data. The fitting coefficients of the exponential decays give: the absorption length, $L_{abs}=1.6$mm, and the losses (i.e., absorption and scattering) length, $L_{los}=1.2$mm, for the pure dye solution and for the $0.03$w/w silica-dye solution, respectively; this implies that the effect of the particles on the losses is very small.

In summary the above analysis shows that the role of disorder is predominantly to introduce random phase modulation. Moreover, since absorption does not qualitatively affect shock formation, the disorder induced phase scrambling is predominant over all the loss mechanisms in determining the shock point $Z_s$ measured below.

\section{Shock point}

In this section we report the procedure to identify the shock point Z$_s$ and to determine the threshold for the wave breaking in terms of laser power and SiO$_2$ concentration.

\begin{figure}[ht]
\includegraphics[width=8.5cm]{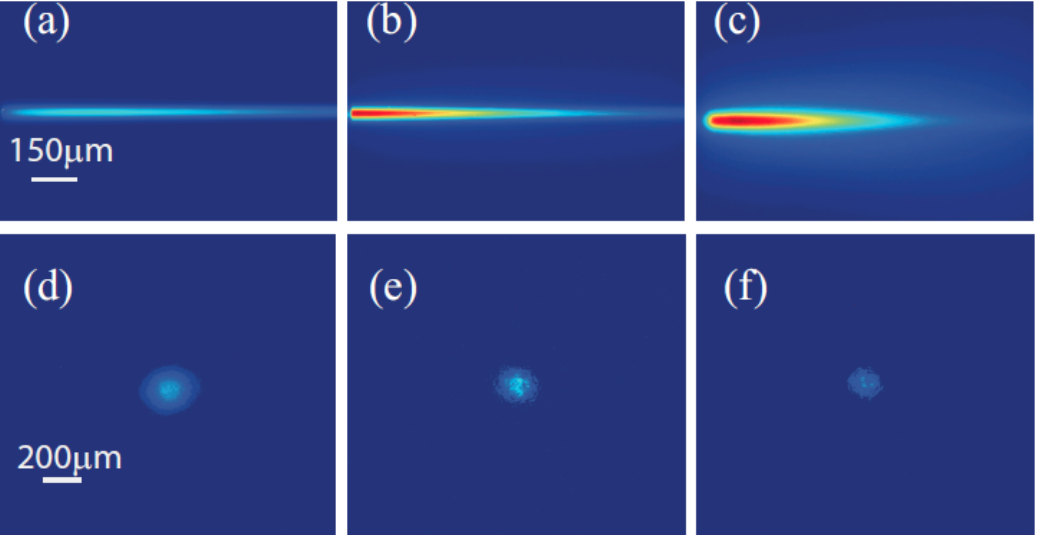}
\caption{(Color online) Top panels: low power ($P=10$mW) images of the fluorescence emission of the beam along $Z$ at different concentrations of silica spheres: (a) $c_{SiO_2}=0$, (b) $c_{SiO_2}=0.017$ w/w, (c) $c_{SiO_2}=0.030$ w/w; bottom panels: corresponding images of the transmitted (in the $X$-$Y$ plane) intensity at the exit facet of $1$mm cell.
\label{figexp4}}
\end{figure}

\begin{figure}[ht]
\includegraphics[width=8.5cm]{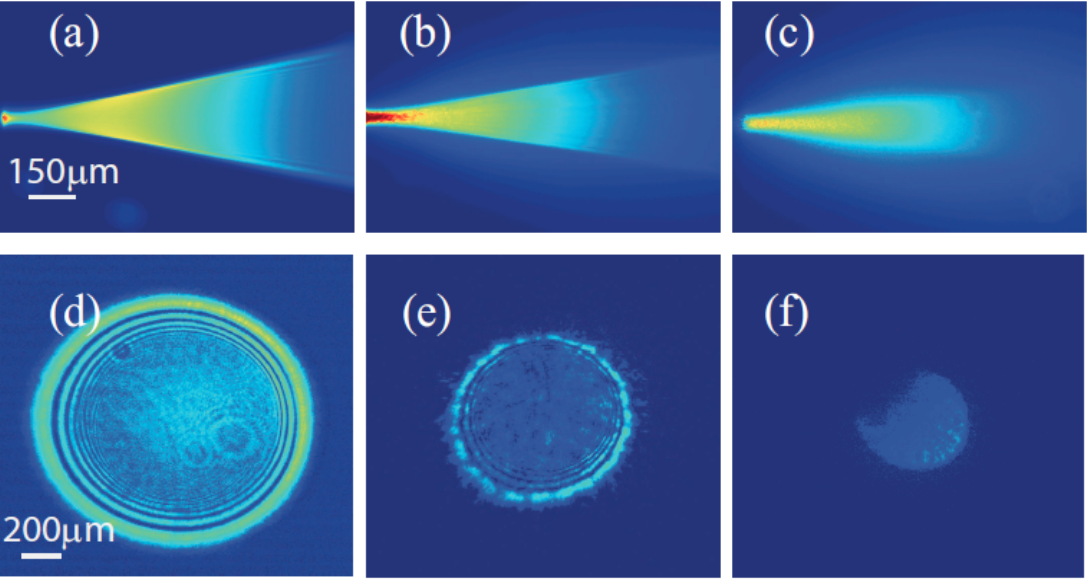}
\caption{(Color online) The same of Fig. \ref{figexp4} at laser power $P=450$mW.}
\label{figexp5}
\end{figure}

In Figs. \ref{figexp4}(a)-\ref{figexp4}(c) and \ref{figexp5}(a) and \ref{figexp5}(c), we show the images of the propagating beam versus Z direction at low and high laser power, respectively. At low power, i.e., $P=10$mW, no nonlinear effects are visible. In Figs. \ref{figexp4}(d)-\ref{figexp4}(f) and \ref{figexp4}(d)-\ref{figexp4}(f) we show the corresponding images of the output intensity field.

\begin{figure}[ht]\includegraphics[width=8.5cm]{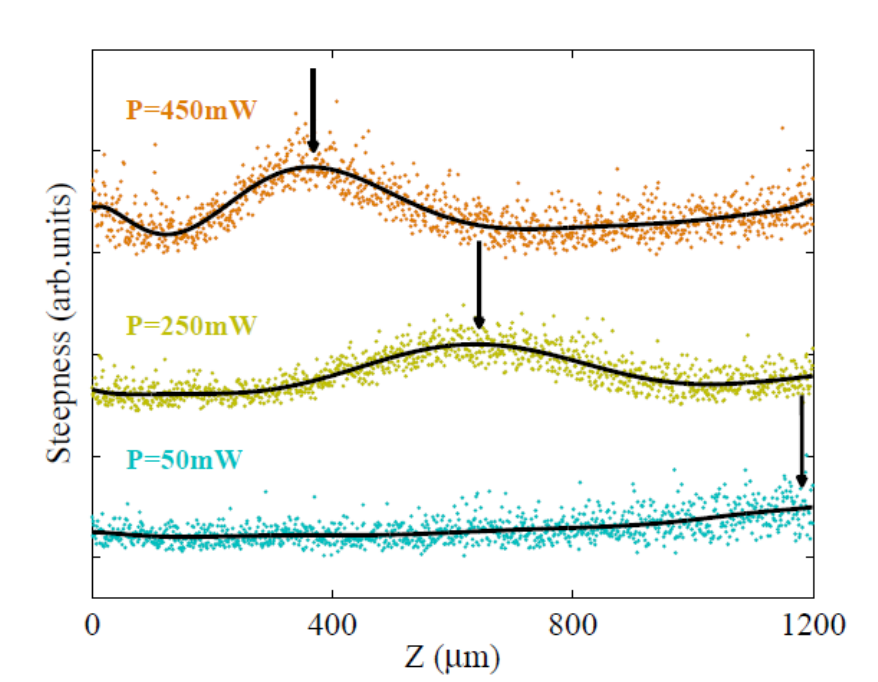}
\caption{(Color online) scattered dots are the calculated steepness along the $Z$ direction for three different powers. The solid lines are polynomial fit of the steepness curves to identify their maximum value, indicated by the arrows.
\label{figexp6}}
\end{figure}

\begin{figure}[ht]\includegraphics[width=8.5cm]{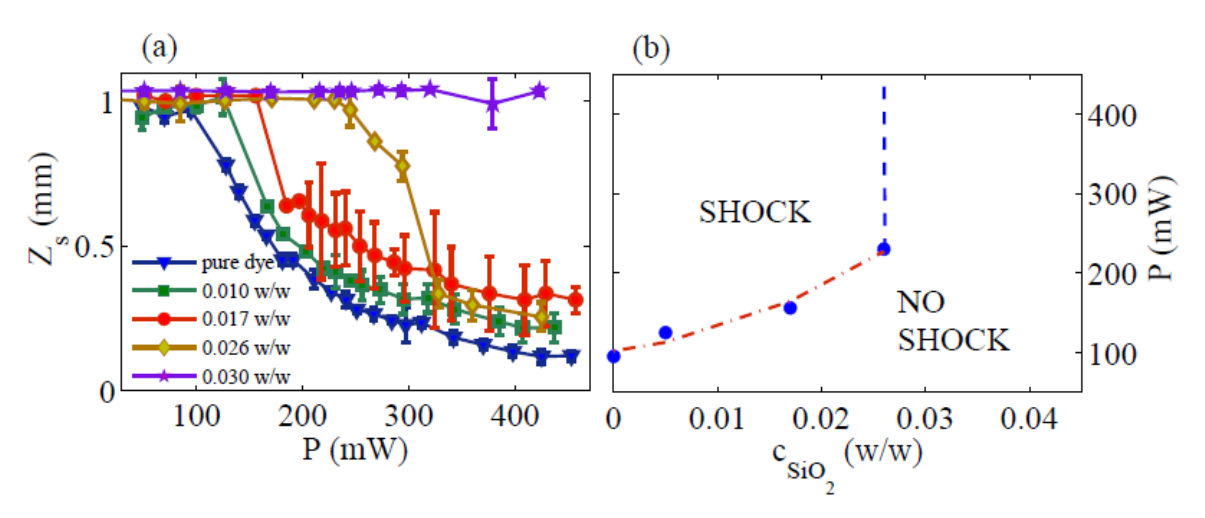}
\caption{(Color online) (a) measured $Z_s$ vs $P$ for different $c_{SiO_2}$; (b) power-disorder diagram from the propagation measurements: filled circles are the threshold power calculated from (a), dashed line is a boundary due to the experimental available observation window, and the dot-dashed line is the boundary as estimated by the theory.
\label{figexp7}}
\end{figure}
Conversely at higher beam power both the effects of nonlinearity and of disorder are evident simultaneously. The two effects are competing as evident by the shock features, i.e., an augmented beam diffraction and the appearance of the undular bores, are enhanced by the laser power and inhibited by the SiO$_2$ concentration, clear from the transverse and longitudinal intensity profiles of Fig. \ref{figexp5}.

We determine the shock point $Z_s$ from the intensity profile of Fig. \ref{figexp4} and \ref{figexp5}, recalling that the shock is originated from a singularity in the phase chirp $|d\phi/dX|\rightarrow\infty$ \cite{ghofraniha,Gho12,Ghoprl12}.

To retrieve the phase singularity from the intensity profile we used the following argument: in the hydrodynamical approximation the laser beam is mainly affected by the defocusing nonlinearity, in a regime of negligible losses and diffraction. Hence at first approximation the phase is proportional to the refractive index perturbation, which in turn depends on the intensity profile because of the Kerr nonlinearity:
\begin{equation}
\phi(X,Y,Z)=\frac{k_0Z}{n_0}\Delta n[I(X,Y,Z)].
\label{eq:phase}\end{equation}
From Eq. (\ref{eq:phase}) we can estimate the occurrence of the singularity in the phase from the intensity profiles, in fact:
\begin{equation}
\nabla_{X,Y}\phi(X,Y,Z)\propto\nabla_{X,Y}I(X,Y,Z).
\label{eq:shockpoint}
\end{equation}
Equation (\ref{eq:shockpoint}) shows that the point of maximum phase chirp is given by the maximum derivative in the intensity profile, this allows the estimation of the shock point as follows: we calculate the transverse derivative of the intensity normalized to the peak value, $I_N$, and we define the steepness $S(Z)$ as the maximum with respect to the transverse coordinates of such a derivative:
\begin{equation}
S(Z)=\text{max}_{X,Y}[\nabla_{X,Y}I_N(X,Y,Z)].
\label{eq:stepness}
\end{equation}
The shock point, $Z_s$, is finally defined as the position of the maximum steepness versus $Z$:
\begin{equation}
Z_S=\text{max}_Z[S(Z)].
\label{eq:Zs}
\end{equation}

In Fig. \ref{figexp6} we show the steepness curves $S(Z)$ at three different laser power $P$. The point of shock occurs at propagation distances consistently smaller with the increase of the incident laser beam power, a signature that, for a fixed level of disorder, the increase of the nonlinearity enhances the shock formation. We note that the curve corresponding to the lowest $P$ shows a monotonous trend and reaches its maximum value at the edge of the observation window. This implies the existence of a threshold value of $P$ below which $Z_s$ assumes a constant value (equal to the size of the observation window $L_0\sim1$mm).

In Fig. \ref{figexp7}(a) we plot the calculated $Z_s$ vs $P$ for all the prepared $c_{SiO_2}$ concentrations. We observe that the threshold power at which $Z_s$ starts to decrease with respect to $L_0$ becomes larger when increasing $c_{SiO_2}$, resulting in a shift of the power threshold towards higher values. In Fig. \ref{figexp7}(b) we map the threshold $P$ in a disorder-power shock phase diagram.

We remark that the obtained $Z_s$ values are in all the investigated cases always smaller of the absorption length $L_{abs}$, confirming that the absorption only marginally affects the shock formation that is instead connected to the phase scrambling
due to the silica particles scattering.

\section{Intensity correlation at the shock}

\begin{figure}[ht]
\includegraphics[width=8.5cm]{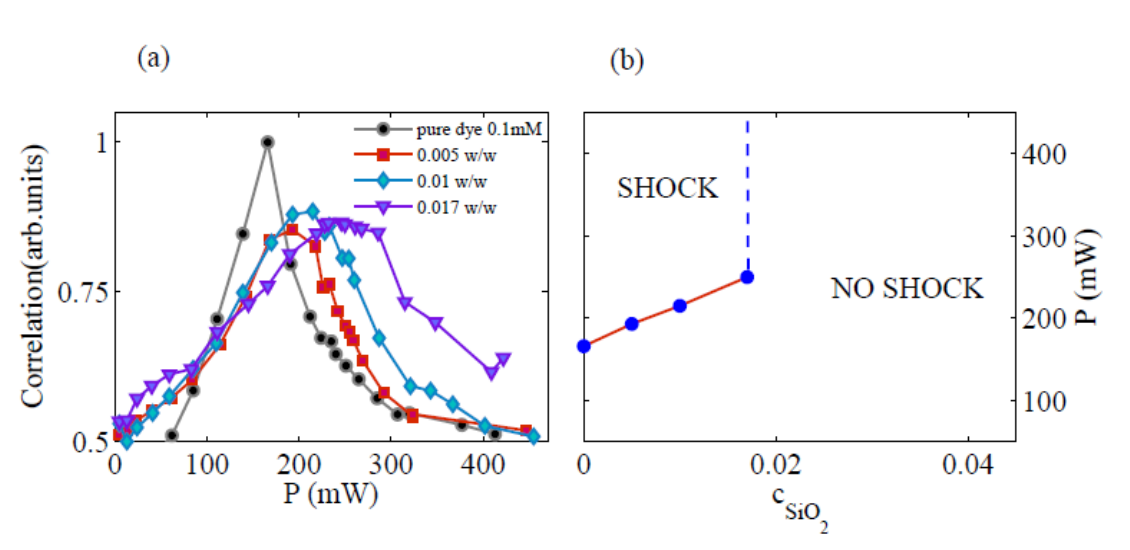}
\caption{(Color online) (a) Correlation curves calculated from the far-field measurements calculated for the ordered sample and for three different increasing values of $c_{SiO_2}$; (b) power-disorder phase diagram as calculated from the transverse intensity correlation curves of (a).
\label{figexp8}}
\end{figure}
In the previous section we have quantitatively analyzed the top fluorescence near- field images of the propagating beam by calculating the shock position $Z_s$. In order to also analyze the transmitted profiles (bottom panels of Figs. \ref{figexp5} and \ref{figexp6}) we calculate the correlation function as follows:
\begin{equation}
C(P)=\Sigma_{i,j}I_P(i,j)I_0(i,j)/\Sigma_{i,j}I_0(i,j)I_0(i,j),	
\label{eq.correlation}\end{equation}
where $P$ is the laser power, $I_P(i,j)$ the intensity distribution on the CCD camera, with $i$ and $j$ pixel indexes,
corresponding to a certain power $P$, and $I_0(i,j)$ is the reference image of the intensity distribution transmitted
from the pure dye sample ($c_{RhB}=0.1$mM) at the laser power $P=160$mW, such reference image was selected as the first
image clearly showing the post-shock rings. The function $C(P)$ provides an estimation of the degree of coherence after
propagation in the scattering samples.

Figure \ref{figexp8}(a) shows the correlation curves $C(P)$ calculated from the bottom images of Figs. \ref{figexp5} and \ref{figexp6}, we observe that the curves grow up to a maximum value and then they start to decrease with the increasing power $P$. The power $P$ of the peak value increases with the SiO$_2$ concentration, meaning that in the presence of disorder a stronger nonlinearity is necessary to overcome the dephasing effect due to the scattering with the silica particles. Figure \ref{figexp8}(b) shows the disorder-power shock phase diagram as calculated from the curves of Fig. \ref{figexp8}(a): the filled circles represent the threshold power $P$, defined as the power at which the maximum correlation between the ordered an the disordered samples is achieved. We stress that the shock phase diagram of Fig. \ref{figexp8}(b) is in good agreement with that of Fig. \ref{figexp7}(b) calculated from the $Z_s(P)$ curves; the slight discrepancy between the two phase diagrams derives from the different definition of the shock point.

\section{Shock threshold from angular spreading measurements}
\begin{figure*}
\includegraphics[width=\textwidth]{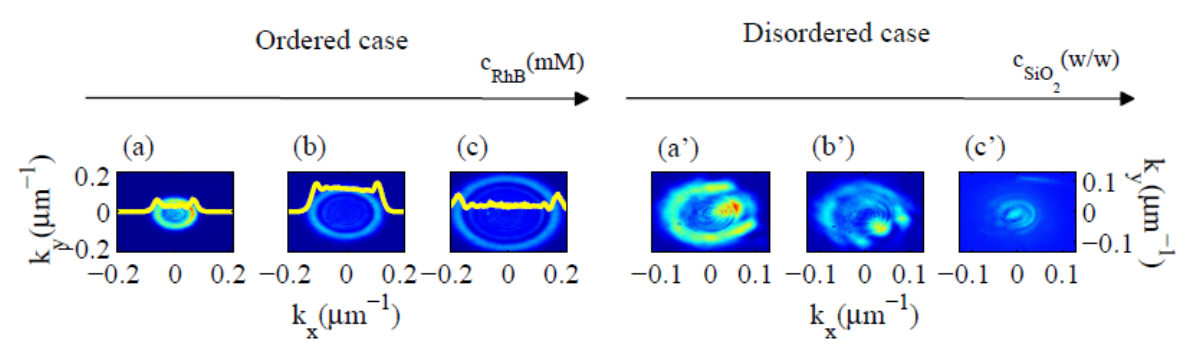}
\caption{(Color online) Images of the far-field intensity distribution of the transmitted beam after $1$mm  propagation distance. The left panels group refers to ordered samples (i.e. fixed $c_{SiO_2}=0$) at fixed power $P=140$mW at various dye concentrations (a) $c_{RhB}=0.05$mM, (b) $0.1$mM (b), (c) and $0.2$mM. The right panels group shows the far field intensity profile in disordered samples (i.e., fixed $c_{RhB}=0.1$mM) fixed laser power $P=130$mW and at various SiO$_2$ concentrations (a') $c_{SiO_2}=0.005$w/w, (b') $0.017$w/w and (c') $0.038$w/w.
\label{figexp9}}
\end{figure*}
The characteristic post-shock annular structure and the diffraction enhancement displayed by the near-field transverse and longitudinal intensity distribution, reveal a non-trivial involvement of the wave-vector spectrum in the shock phenomenon. In this section we report the investigation on the far-field intensity distribution of the transmitted beam after $1$mm propagation distance. Such an investigation allows us to measure the angular aperture $\theta$.
Fig. \ref{figexp9} provides a qualitative overview of the whole set of the far-field measurements. The panels on the left side [Figs. \ref{figexp9}(a)--\ref{figexp9}(c)] report images relative to the ordered samples ($c_{SiO_2}=0$) at fixed power $P=130$mW and at various dye concentrations $c_{RhB}$ ranging from $0.05$ to $0.2$mM; the right panels [Figs. \ref{figexp9}(a')--\ref{figexp9}(c')] refer to the disordered samples at fixed power $P=140$mW prepared at $c_{RhB}=0.1$mM and varying $c_{SiO_2}$ between $0.005$w/w and $0.038$w/w. The way the nonlinearity and the disorder affect the shock phenomenology, i.e., the appearance of the characteristic rings and the enlargement of the spectral content, reveals that their effect on the shock formation is opposite: the images [Figs.\ref{figexp9}(a)--\ref{figexp9}(c)] show an enhancement with the increase of $c_{RhB}$ (i.e., of the strength of nonlinearity); conversely those in Figs. \ref{figexp9}(a')--\ref{figexp9}(c') show the inhibition of shock with $c_{SiO_2}$.

Note that in images on the right the circular symmetry of the DSWs is lost because of the refractive index inhomogeneities. In other words, the shock wave has a partially randomized spatial distribution. In order to quantitatively analyze both the sets of measurements, we perform a radial average of the two-dimensional collected profiles and we estimate the angular aperture $\theta$ as the full width half maximum since the profile appears as a single peak; and as the distance between the two leading peaks when the profiles start to split because of the wave breaking due to the defocusing nonlinearity.

In what follows we detail the results obtained for the ordered and disordered case \cite{Gentilini12}.

\subsection{Ordered case}
\begin{figure}[ht]
\includegraphics[width=8.5cm]{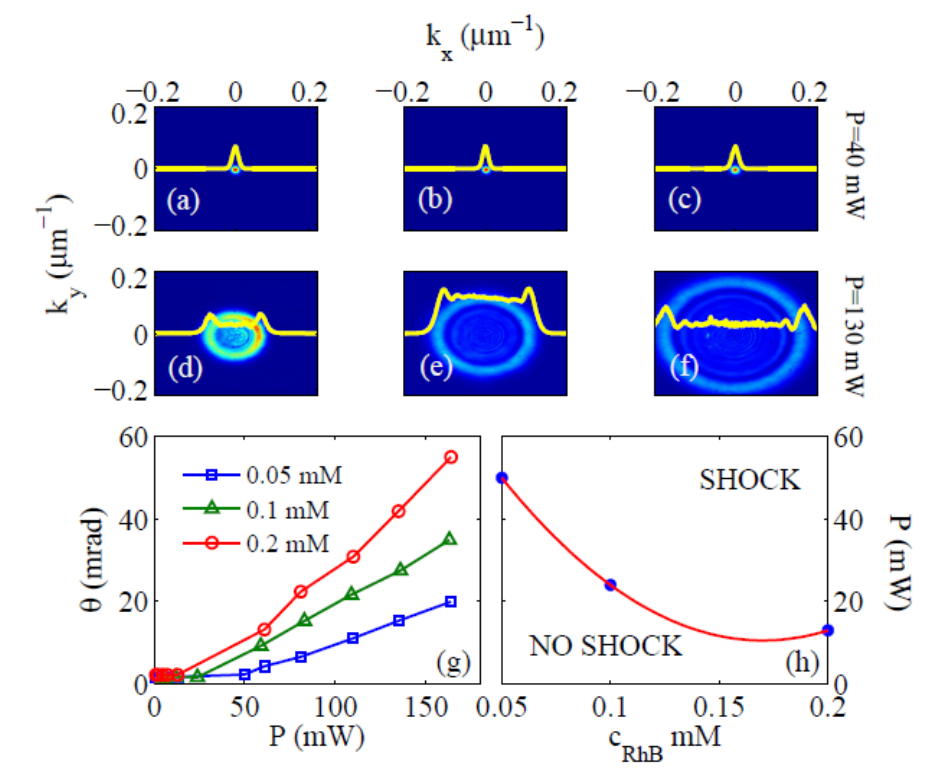}
\caption{(Color online) (a)--(f): spectral content of the transmitted beam of the ordered samples ($c_{SiO_2}=0$) at  two laser power $P$ and different dye concentration: (a),(d) $c_{RhB}=0.05$mM; (b),(e) $0.1$mM and, (c), (f) $0.2$mM. Bottom panels: $\theta$ vs $P$ curves (g); threshold power $P$ as calculated from curves of (g) vs $c_{RhB}$(h).
\label{figexp10}}
\end{figure}
We study the occurrence of DSW in the pure dye solutions ($c_{SiO_2}=0$) when varying input laser power $P$ for different dye concentrations $c_{RhB}$.

 Figures \ref{figexp10}(a)-\ref{figexp10}(f) display the collected images of the far-field intensity distribution when a low [Figs. \ref{figexp10}(a)-\ref{figexp10}(b)] and high [Figs. \ref{figexp10}(d)-\ref{figexp10}(f)] power laser beam impinges on the pure dye solutions. We note that the higher the dye concentration, the larger the spatial spectral content due to the higher nonlinearity.

Figure \ref{figexp10}(g) shows the curves of the calculated angular aperture $\theta$ vs $P$ as obtained for the different $c_{RhB}$ concentrations.
In these measurements both the control parameters contribute to strengthen the nonlinearity of the system. Consistently we find that, above a critical power, $\theta$ starts to increase with $P$ because of the speedup of the shock formation due to the augmented nonlinearity; the slope of the curves increases with $c_{RhB}$, providing larger spectra at the same laser power $P$.

Analogously to our analysis of the shock position $Z_s$, we seek also for the angular aperture a threshold value for the laser power. Such a threshold power can be retrieved in the above mentioned power value, beyond which $\theta$ starts to linearly grow with $P$. We plot the threshold power values in the diagram of $P$ versus $c_{RhB}$ in Fig. \ref{figexp10}(h).

\subsection{Disordered case}
\begin{figure}[ht]
\includegraphics[width=8.5cm]{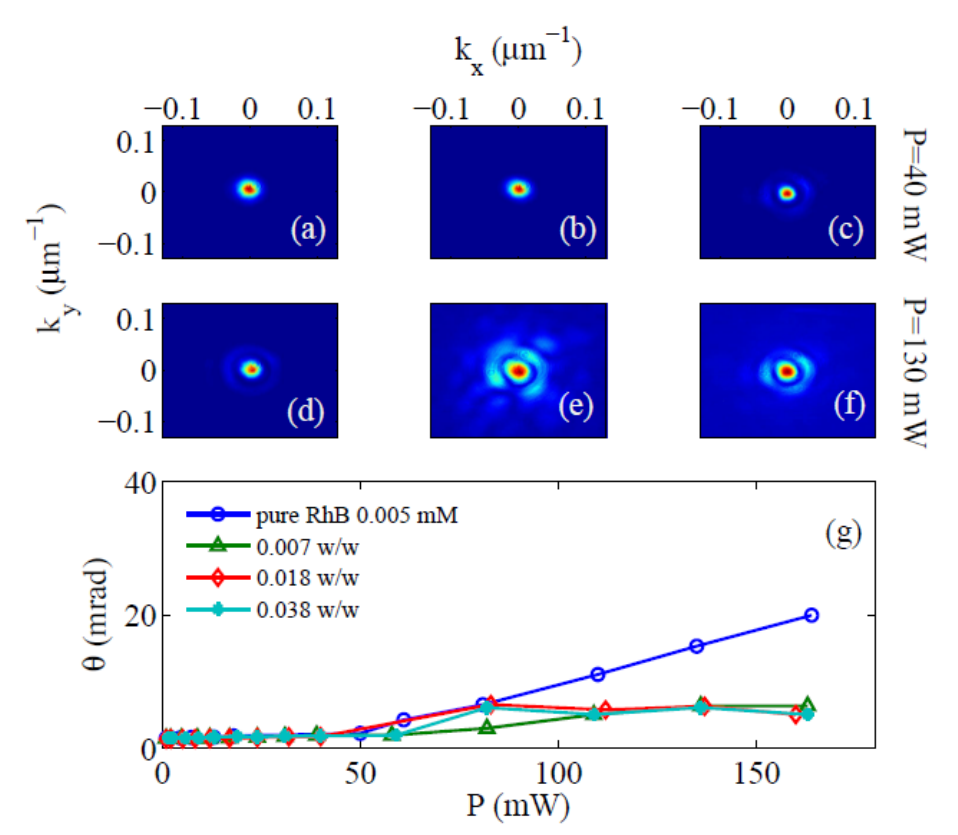}
\caption{(Color online) (a)-(f): low dye concentration ($c_{RhB}=0.05$mM) spectral content of the transmitted beam by disordered samples at two different laser power $P$ and different SiO$_2$ concentration: (a), (d) $c_{SiO_2}=0.007$w/w; (b), (e) $0.018$w/w and
(c), (f) $0.038$w/w; (g) the angular aperture $\theta$ vs $P$: the curves show no threshold behavior in the presence of disorder.
\label{figexp11}}
\end{figure}

We consider the interplay between disorder and nonlinearity in the DSW formation by dispersing the SiO$_2$ particles in pure dye solutions at $c_{RhB}=0.05$mM and $c_{RhB}=0.1$mM. Figures \ref{figexp11}(a)-(f), show the spectral profiles for different $c_{SiO_2}$ and laser power $P$ at fixed $c_{RhB}=0.05$mM. At this dye concentration and any laser power $P$, no shock formation emerges from the spectra as can be retrieved also in the trend of $\theta$ vs $P$ reported in Fig. \ref{figexp11} (g). This is a signature of the fact that at the lowest prepared dye concentration the nonlinearity is counteracted by the disorder which prevents the appearance of any shock phenomenology.

\begin{figure}[ht]
\includegraphics[width=8.5cm]{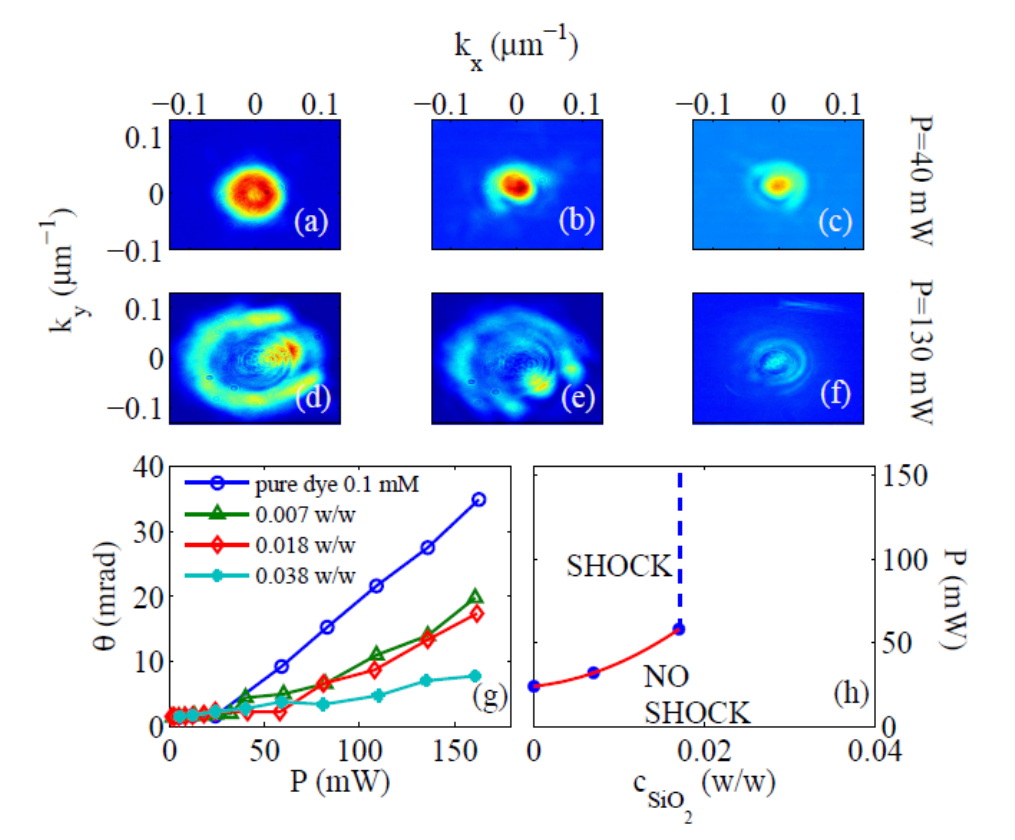}
\caption{(Color online) (a)-(f) are the same as Fig. \ref{figexp11} but at higher dye concentration ($c_{RhB}=0.1$mM);
(g) the angular aperture $\theta$ vs $P$: at this higher dye concentration the curves have recovered the threshold behavior
also in the presence of disorder; (h) power-disorder phase diagram from the curves of (g).
\label{figexp12}}
\end{figure}

In Fig. \ref{figexp12} we show the case of the disordered samples obtained by the $0.1$mM pure dye solution. At the higher power $P=140$mW the shock characteristic rings are clearly visible in the spectra corresponding to the lower concentrations of SiO$_2$, $c_{SiO_2}=0.007$ and $0.018$w/w and disappear at the highest concentration $c_{SiO_2}=0.038$.

In Fig. \ref{figexp12}(g) we show the curves $\theta$ vs $P$ calculated from the images of the upper panels [Figs. \ref{figexp12}(a)-\ref{figexp12}(f)]. We retrieve the expected threshold behavior with respect to both the control parameters $P$ and $c_{SiO_2}$, which results in the power-disorder shock phase diagram of Fig. \ref{figexp12}(h).

\section{Conclusion}
\vspace{-0.5cm}
We have reported a detailed analysis of our experiments aimed at understanding the role of disorder in the occurrence of dispersive shock waves in a thermal defocusing medium. We collected the propagating and the transmitted intensity profile of a CW laser beam impinging on aqueous solutions of Rhodamine B, with an added controllable amount of disorder achieved by dispersing silica beads at well-defined concentrations.
Resorting to the hydrodynamical approximation we analyzed the collected intensity distributions associated to the two observables of the system: the shock point from the propagating intensity profiles and the angular aperture of the transmitted intensity profiles. Both the observables have evidenced the expected thresholds for the occurrence of the shock phenomenon with respect to the degree of nonlinearity and in the amount of disorder.
The calculation of the shock point has in fact led to the first determination of disorder-power shock phase-diagram; also the trend of the angular aperture versus the laser power for the different silica concentrations has allowed to the calculation of two distinct shock diagrams related to the ordered and disordered cases. We also analyzed the degree of correlation of the shock images when increasing disorder.
These experiments open the way to further investigations concerning the interplay between disorder and nonlinearity, with ramifications in several research directions, from basics physics, as the study of nonlinear waves in random media, to applied research, where the exploitation of nonlinear effects in disordered media, such as biological tissue and atmosphere, should be fundamental in order to improve spectroscopy and imaging.
\section{acknowledgments}
The research leading to these results has received funding from the European Research Council under the European Community's Seventh Framework Program (FP7/2007-2013)/ERC Grant No. 201766, from the Italian Ministry of Research (MIUR) through the PRIN Project No. 2009P3K72Z, and from  the Italian Ministry of Education, University
and Research under the Basic Research Investigation Fund (FIRB/2008) program/CINECA Grants No.
RBFR08M3P4 and No. RBFR08E7VA. We thank M. Deen Islam for technical assistance.

\end{document}